\documentclass[nofootinbib,showpacs,showkeys]{revtex4}
\usepackage{graphicx,color}
\usepackage{amssymb}
\usepackage{amsmath}
\usepackage{bm}
\usepackage{lipsum}
\usepackage{latexsym}
\usepackage{dcolumn}
\usepackage{float}
\usepackage{hyperref}

\begin{document}

\title{Cosmological inviability of $f(R,T)$ gravity}

\author{Hermano Velten}\email{velten@pq.cnpq.br}
\author{Thiago R. P. Caram\^es}\email{thiago.carames@ufes.br}

\affiliation{Universidade Federal do Esp\'{\i}rito Santo (UFES), Av. Fernando Ferrari S/N, 29075-910, Vit\'oria, Brazil}

\begin{abstract}

Among many alternative gravitational theories to General Relativity (GR), $f(R,T)$ gravity (where $R$ is the Ricci scalar and $T$ the trace of the energy-momentum tensor) has been widely studied recently. By adding a matter contribution to the gravitational Lagrangian, $f(R,T)$ theories have become an interesting extension to GR displaying a broad phenomenology in astrophysics and cosmology. In this paper, we discuss however the difficulties appearing in explaining a viable and realistic cosmology within the $f(R,T)$ class of theories. Our results challenge the viability of $f(R,T)$ as an alternative modification of gravity.

\end{abstract}

\maketitle

\section{Introduction}

Astronomical observations during the 20th century unveiled two of the biggest phenomena in the universe: dark matter and dark energy. The former behaves as an invisible dust matter favoring the process of gravitational clustering while the latter yields a late time acceleration of the cosmological background. It is however worth noting that such intriguing mechanisms (or substances) have emerged complementing the pure general relativity (GR) based description of gravity.

By avoiding the insertion of new unknown ingredients the search for alternative theories to GR has become a fruitful investigation route. Since the GR can be modified in many distinct directions the consequence is the large number of theories available in the literature \cite{MG}.

What is the best alternative theory? There is no clear answer to this question. However, the set of available options has been reduced over the past years. Gravitational theories can be ruled out if they do not successfully describe modern astronomical observations or because of theoretical incompatibilities and unphysical features, e.g., the appearance of ghosts.

Our aim in this work is to show the inviability of one of such proposals when applied to the cosmological context. We focus on the so-called $f(R,T)$ theory where $R$ is the Ricci scalar and $T$ is the trace of the stress-energy tensor. The GR case of such theory corresponds to the standard Einstein-Hilbert term $f(R,T)=R$ in the gravitational Lagrangian. This model has been introduced in Ref. \cite{Harko:2011kv} as a possible interpretation for a running cosmological constant. Then, one expects that this modification of gravity accounts for the accelerated expansion. Dark matter seems therefore to be a fundamental ingredient in the $f(R,T)$ scenario. 

It is worth noting that the mapping between well known cosmological scenarios (e.g., the $\Lambda$CDM one) have being reconstructed from the $f(R,T)$ theory as for example in Ref. \cite{Momeni:2011am}. This reference also claims an agreement of the theory with the Baryonic Acoustic Oscillation (BAO) data, but a proper confrontation between theory and data is not performed.  

We show that most of the available $f(R,T)$ models in the literature do not lead to a viable cosmological background expansion. We use current available expansion data to show that in the low-redshift domain there is no compatible cosmological expansion in the $f(R,T)$ gravity. Some cases do not even lead to a late time accelerated expansion which is a well established observational fact \cite{Seikel:2008ms}. 

The indication that $f(R,T)$ theories have difficulties in describing the linear growth of matter perturbations has been already discussed in Ref. \cite{alvarenga}. This conclusion was made however on a qualitative ground without proper comparison with observation data like the Cosmic Microwave Background spectrum, Baryonic Acoustic Oscilations measurements or the matter power spectrum $P(k)$.

In the next section we review the $f(R,T)$ gravity applying it to a Friedmann-Lemaitre-Robertson-Walker (FLRW) metric. We discuss how realistic cosmologies are built in the realm of $f(R,T)$ gravity pointing out some typical misunderstandings in the literature.

It is convenient to bring to the reader's attention an important assumption we are going to make throughout this paper. For the Riemman tensor we adopt the standard convention $R_{\ \beta\gamma\delta}^{\alpha}=\partial_\gamma\Gamma_{\ \beta\delta}^\alpha-\partial_\delta\Gamma_{\ \beta\gamma}^\alpha+\Gamma^{\alpha}_{\ \mu\gamma}\Gamma^{\mu}_{\ \beta\delta}-\Gamma^{\alpha}_{\ \mu\delta}\Gamma^{\mu}_{\ \beta\gamma}$, from which the Ricci tensor is obtained by contracting the first index with the third one, e.g $R_{\alpha\beta}= R_{\alpha\gamma\beta}^\gamma$. Such choice is the most used by some of the main GR textbooks \cite{misner,wald,schutz,dinverno} and is widespreadly adopted in the literature. So, we believe it is useful to emphasize this point in order to help some readers to not be misled by less usual definitions as for instance that one made by the authors in \cite{alvarenga}, who considers the Weinberg's assumption \cite{weinberg}. In the latter case the Riemann tensor appears with opposite sign to that one adopted in this work, which coincides with the conventions used in \cite{Harko:2011kv}. Besides, it can be checked that this choice shall influence the sign of the term involving the differential operator $g_{\mu\nu} \Box-\nabla_{\mu}\nabla_{\nu}$. In the particular case of the reference \cite{alvarenga}, the authors adjust such sign difference through a proper definition for the energy-momentum tensor.   

In the section III we list four cases of $f(R,T)$ theories which comprise most of the ones investigated in the literature. For the first three examples we compute the respective deceleration parameters and confront these models with the observational data for $H(z)$ (inferred from cosmic chronometers) and Supernovae type $Ia$ data. In particular, we complement the model III analysis (see below) with Supernovae data constraints. These model are plagued due to the fact they do not lead to accelerated phases pf the universe. According to available literature \cite{Moraes:2016jyi} our model IV is in principle viable and promotes the transition from deceleration to acceleration. However, on the model IV we point out our disagreement with the authors of \cite{Moraes:2016jyi}, showing explicitly that their conclusions about the deceleration-acceleration transition is based on an erroneous deceleration parameter (we make this issues clear in the appendix section). The section IV brings an alternative approach for the $f(R,T)$ theory, in which such modified gravity is assumed as the underlying theoretical scenario for the dark matter hypothesis, whereas the usual cosmological constant remains responsible for the cosmic acceleration. The final section is dedicated to the discussion of our results and the concluding remarks of this work. 

\section{The background cosmological dynamics in f(R,T) gravity}

In an $f(R,T)$ theory the gravitational action is given by
\begin{equation}
S=S_G+S_m=\frac{1}{2 \kappa^2}\int d^4 x \sqrt{-g} f(R,T)+\int d^4 x \sqrt{-g} \mathcal{L}_m
\end{equation}
where $\kappa^2=8\pi G$, $T = T^{\mu}_{\mu}$ and $\mathcal{L}_m$ is the Lagrangian density of the matter fields. We adopt $c=1$.
The energy-momentum tensor is defined in terms of the matter action as follows
\begin{equation}
\label{emt}
T_{\mu\nu}= -\frac{2}{\sqrt{-g}}\frac{\delta S_m}{\delta g^{\mu\nu}},
\end{equation}
which by turn leads to
\begin{equation}
\label{emt1} 
T_{\mu\nu}=g_{\mu\nu}{\cal L}_m-2\frac{\partial {\cal L}_m}{\partial g^{\mu\nu}}.
\end{equation}
In the metric formalism, one considers the metric tensor as the only dynamical variable of the respective gravitational theory, so that the corresponding field equations are obtained by varying the action with respect to the metric. Following this procedure we get
\begin{eqnarray}
\label{feqs}
f_R(R,T)R_{\mu\nu}-\frac{1}{2}f(R,T)g_{\mu\nu}+(g_{\mu\nu} \Box-\nabla_{\mu}\nabla_{\nu})f_R \left(R,T\right) = \left[\kappa^2 -f_T(R,T)\right]T_{\mu\nu}-f_T \Theta_{\mu\nu},
\end{eqnarray}
where
\begin{equation}
\label{theta}
\Theta_{\mu\nu}\equiv g^{\alpha \beta} \frac{T_{\alpha\beta}}{\delta g^{\mu\nu}}=-2T_{\mu\nu}+g_{\mu\nu}\mathcal{L}_m-2 g^{\alpha\beta}\frac{\partial^2 \mathcal{L}_m}{\partial g^{\mu\nu}\partial g^{\alpha\beta}}.
\end{equation}
We assume the standard cosmological scenario in which the universe behaves as a homogeneous and isotropic fluid at large scales. This hypothesis is envisaged by a Friedmann-Lema\^itre-Robertson-Walker spacetime
\begin{equation}
\label{flrw}
ds^2 = dt^2 - a(t)^2 \delta_{ij}dx^i dx^j\ ,
\end{equation}
along with an energy-momentum tensor of a perfect fluid written in terms of its energy density $\rho$ and the pressure $p$,
\begin{equation}
\label{pf}
T_{\mu\nu}=(\rho+p)u_{\mu}u_{\nu}-p g_{\mu\nu}.
\end{equation}
Comparing (\ref{emt1}) and (\ref{pf}) we can write the Lagrangian as $\mathcal{L}_m=-p$, while the tensor (\ref{theta}) reduces to $\Theta_{\mu\nu}=-2T_{\mu\nu}-pg_{\mu\nu}$. In this work we shall focus on a classe of $f(R,T)$ theories where the generalized Einstein-Hilbert Lagrangian is given by $f(R,T)=f_1(R)+f_2(T)$, with $f_1(R)$ and $f_2(T)$ being functions purely dependent upon $R$ and $T$, respectively. 
Thus, the modified field equations are
\begin{equation}
\label{eq00}
-3(\dot{H}+H^2)f'_{1}-\frac{f_{1}}{2}-\frac{f_{2}}{2}+3H\dot{f'_{1}}=\kappa^2 \rho+f'_{2}(1+w)
\end{equation}
and
\begin{equation}
\label{eq11}
(\dot{H}+3H^2)f'_{1}+\frac{f_{1}}{2}+\frac{f_{2}}{2}-\ddot{f'_{1}}-2H\dot{f'_{1}}=\kappa^2 p, 
\end{equation}
where the prime and dot denote derivatives with respect to the argument, i.e., $f^{\prime}_1= d\, f_1(R) / dR$, and to the cosmic time, respectively. 

A general $f(R,T)$ model gives rise to a deviation from the usual conservation law, implying in the following form for the continuity equation 
\begin{equation}
\label{consfrt}
\dot{\rho} +3 H\rho(1+w)= -\frac{1}{\kappa^2+f'_{2}}\left[(1+w)\rho \dot{f'_{2}}+ w \dot{\rho} f'_{2}+\frac{1}{2}\dot{f}_{2}\right],
\end{equation}
where $w$ is the parameter of the equation of state, which for a barotropic fluid is given by $p=w\rho$. Notice that (\ref{eq00}), (\ref{eq11}) and (\ref{consfrt}) form a system of three independent differential equations. The fourth-order derivatives of the metric appearing in these equations, $\ddot{f'_{1}}$, enhances the arising of new degrees of freedom in $f(R,T)$ theory, so that along with the variables $a$ and $\rho$, it is also necessary to consider $\ddot{a}$ as an independent variable in order to provide a solution for such system of equations.

The covariant conservation of $T_{\mu\nu}$ is an essential feature of GR, which manifests as an immediate consequence of the diffeomorphism invariance of the theory. So, it is expected that any classical gravitational theory shall in principle obey such requirement as well. 
In the context of interacting dark energy models, a local violation of $\nabla_{\mu}T^{\mu\nu}=0$ may be allowed by means of a possible exchange of either energy or momentum (or both) between the two dark components. Nonetheless, even in these models this exchange occurs in such a way as to preserve the conservation of the total dark fluid. Differently, the equation (\ref{consfrt}) shows a non-conservation of the matter-energy content as a whole, revealing a significant drawback of this class of $f(R,T)$ theories. 
In light of this, in \cite{alvarenga} the authors imposed by hand the fulfillment of (\ref{consfrt}) by setting to zero the expression inside the bracket, e.g $(1+w)\rho \dot{f'_{2}}+ w \dot{\rho} f'_{2}+\frac{1}{2}\dot{f}_{2}=0$. By using a chain rule, one can get rid of the time derivatives and write this constraint condition as a second order differential equation for the function $f_2(T)$, whose integration provides a solution in the form
\begin{equation}
\label{sol}
f_2(T)=\sigma T^{\frac{3w+1}{2(w+1)}}+\sigma_0,
\end{equation}
where $\sigma$ and $\sigma_0$ are integration constants. We may avoid the trivial case $f_2(T)=\textrm{const.}$ by assuming the necessary condition $\omega\neq - 1/3$. Also, $\omega\neq + 1/3$ is adopted in order to assure that $T \neq 0$. Considering a dustlike matter, for which $\omega=0$, the solution above becomes
\begin{equation}
\label{sol}
f_2(T)=\sigma T^{\frac{1}{2}}+\sigma_0.
\end{equation}
The authors then argued that this model should represent the only viable $f(R,T)$ theory, as it constitutes the only case in which the standard conservation law is preserved, what implies in automatically ruling out anyone else. This result imposes a stringent restriction on the $f(R,T)$ gravity. They used such viable $f(R,T)$ model to study the linear evolution of matter density perturbations within the sub-Hubble regime and assuming a quasi-static approximation\footnote{This means to neglect all the time derivatives of the Bardeen potentials present in the perturbative equations.}. However, they found an inconvenient scale dependence for the dynamics of the perturbations which would strongly disagree with expected results, but not properly comparing such prediction with data. Nevertheless, Ref. \cite{alvarenga} already brings the message that $f(R,T)$ theories are disfavored. 

Our purpose in this work aims to reinforce at background level the difficulties these $f(R,T)$ theories showed to experience at perturbative regime, which can undermine them as a viable way out to explain the observed universe. 

\section{The usual choices for $f(R,T)$}
As mentioned in the previous section, we will concentrate our attention on the $f(R,T)$ theories obeying the form 
\begin{equation}\label{fRT0}
f(R,T) = f_1(R) + f_2(T).
\end{equation}\label{sol}
The simplest and trivial choice for the $R$ dependence corresponds to the Einstein-Hilbert $f_1(R)=R$ term. This is the way to study how the material corrections given by $f_2(T)$ promote deviations from GR. With he following choice
\begin{equation}\label{fRT}
f(R,T)= R + \lambda T + \gamma_{n} T^n,
\end{equation}
we can cover most of the proposed $f_2(T)$ functions in the literature. The parameters $\lambda$ and $\gamma_{n}$ are arbitrary constants. The subscript appearing in $\gamma_n$ refers to the respective powers of the term $T^{n}$ in (\ref{fRT}). 

Given $f_1(R)=R$, let us recall that the departure from GR encoded in (\ref{fRT0}) means a minimal coupling between the curvature and the energy-momentum tensor, so that the presence of matter fields constitutes an essential condition for the effects of the modification of gravity to be perceived. So, there are two possible ways to recover GR from (\ref{fRT}): setting the coefficients $\lambda$ and $\gamma_n$ to zero or merely considering the vacuum case, $T=0$. But, notice however that the latter case is degenerated with the choice $\omega= 1/3$. Besides, it is easy to notice that the GR plus cosmological constant can be obtained by fixing $n=0$ along with $\lambda=0$ and $\gamma_0=2\Lambda$. So, in order to ensure that $f(R,T)$ behaves as a regular function at the vacuum limit (which as mentioned above shall coincide
with GR vacuum), it is reasonable to assume positiveness for the powers of $T$, i.e., $n\geq 0$.

Given the barotropic equation of state $p=w\rho$ the background expansion in a FLRW universe can be written in terms of the fractionary density $\Omega= \rho / \rho_{c0}$, where the critical density reads $\rho_{c0}=3H^2_0/\kappa^2$,
\begin{equation}\label{H1}
\frac{H^2}{H_0^2} = \Omega + \frac{3}{2}\bar{\lambda}\left(1-\frac{w}{3}\right)\Omega+\bar{\gamma_n}\left[(1+w)n(1-3w)^{n-1}+\frac{(1-3w)^{n}}{2}\right]\Omega^{n},
\end{equation}
where we have defined the constants $\lambda$ and $\gamma_n$, in terms of new dimensionless parameters
\begin{eqnarray}
\label{params}
\bar{\lambda} = \frac{\lambda}{\kappa^2}\ ; \hspace{2cm}   \bar{\gamma}_n=\frac{\gamma_n(3 H^2_0)^{n-1}}{\kappa^{2n}}\ .
\end{eqnarray}
In the standard cosmology one considers pressureless matter $p=0$, then $\Omega = \Omega_0 a^{-3} $ where the scale factor today is set to the unity $a_0=1$. In a flat $\Lambda$CDM cosmology $\Omega_{0}=0.3$ and $\Omega_{\Lambda}=0.7$.

It is barely noticed in the literature that expansion (\ref{H1}) imposes the following constraint on $\Omega(z=0)=\Omega_{0}$, 
\begin{equation}\label{H0constraint}
1 = \Omega_0 + \frac{3\bar{\lambda}}{2}\left(1-\frac{w}{3}\right)\Omega_0+\bar{\gamma}_n\Omega^n_0 \left[(1+w)n(1-3w)^{n-1}+\frac{(1-3w)^n}{2}\right].
\end{equation}
Therefore, the usual interpretation of the parameter $\Omega_0$ as being a fraction of the today's critical density of the fluid does not trivially apply when $\bar{\lambda}\neq 0$ or $\bar{\gamma}_n \neq 0$ though since energy density is a positive definite quantity the requirement $\Omega_0 >0$ imposes lower bounds on the $\bar{\lambda}$ and $\bar{\gamma}_n$ values.

Since $f(R,T)$ theories are supposed to explain the accelerated expansion i.e., the dark energy phenomena, we assume for simplicity that $\rho$ represents the total matter distribution. It is a reasonable approximation to neglect now the $5\%$ contribution in the baryonic sector. 

It is also worth to point out that modified gravity theories can be recasted in the standard GR form such that $R_{\mu\nu} - (1/2) R g_{\mu\nu} = \kappa^2 ( T^m_{\mu\nu} + T^{eff DE}_{\mu\nu})$ where all the new geometrical terms (appearing in the left hand side) are grouped (in the right hand side) to form an effective dark energy contribution $T^{eff DE}_{\mu\nu}$. According to this, Eq. (\ref{H1}) reads $H^2 /H^2_0 = \Omega + \Omega^{eff DE}$. This is a standard procedure in $f(R)$ theories, for example. In $f(R,T)$ however such decomposition is not useful since such effective dark energy density would still be coupled to the actual matter field i.e., $\Omega^{eff DE} \equiv \Omega^{eff DE} (\Omega)$. Therefore, one can not constrain the theory from a bound such that $\Omega^{eff}_{DE} >0 $ since this is degenerated with the $\Omega_0$ value.

In the standard cosmology, the temporal dependence of the density $\rho$ in a FLRW universe is found via the usual conservation law $\nabla_{\mu}T^{\mu\nu}=0 \rightarrow \dot{\rho}+3H(\rho+p)=0$. Rather than, in $f(R,T)$, from (\ref{consfrt})

\begin{equation}\label{conservation}
\dot{\Omega}+3H\Omega(1+w)=\frac{-\dot{\Omega}}{1+\bar{\lambda}+n\bar{\gamma}_n \Omega^{n-1}(1-3w)^{n-1}}\left\{\frac{\bar{\lambda}(1-w)}{2}+n\bar{\gamma}_n \Omega^{n-1}(1-3w)^{n-1}\left[\frac{2n(1+w)-(1+3w)}{2}\right]\right\}.
\end{equation}

With the choice (\ref{fRT}) we identify the following sub models:

\begin{itemize}
\item Model I: $f(R,T)= R + \lambda T$ with ($\gamma_n=0$); 
\item Model II: $f(R,T)= R + \gamma_{2} \,T^2$ with ($\lambda=0$);
\item Model III: $f(R,T)= R + \gamma_{1/2} \,T^{1/2}$ with ($\lambda=0$);
\item Model IV: $f(R,T)= R + \lambda T + \gamma_{2} \,T^{2}$ with ($\lambda \neq 0$ and $\gamma_{2} \neq 0$ ).
\end{itemize}

We proceed comparing the background expansion in $f(R,T)$ models with observational data. We investigate now whether the cosmologies provided by models I, II and III pass the test of describing available background data. For the Figs. I, II and III we use typical $H(z)$ data obtained from the differential evolution of cosmic chronometers (galaxies that are assumed to passively evolve and therefore provide a direct estimate of $H(z)=-1/(1+z) dz/dt \cong -1/(1+z) \Delta z / \delta t$) \cite{Hz}. The data points shown in these figures consist of 28 data points listed in \cite{Hzdata}.

Model I is the simplest version of $f(R,T)$ gravity and the most common one in the literature. It is worth noting that the energy density is a positive definite quantity and therefore $\Omega(z) > 0$, $\forall z$. Therefore, from (\ref{H0constraint}) one sets the bound
\begin{equation} \label{H0constraintmodelI}
1 = \Omega_0 \left[1+\frac{3\bar{\lambda}}{2}\left(1-\frac{w}{3}\right)\right]\hspace{1cm} \rightarrow \hspace{1cm}\bar{\lambda} \left(1-\frac{w}{3}\right) > -\frac{2}{3}.
\end{equation}
For the pressureless case, as expected in a pure CDM scenario, $\bar{\lambda} > -2/3$.

Left (right) panel of Fig. 1 shows the evolution of the expansion rate with the redshift for model I (II). The inset in this figure amplifies the low-z region. The $H(z)$ data used here lies in the range $0 < z < 2$. All curves assume the reference value $H_0=70 km/s/Mpc$ and pressureless matter fluid $w=0$. Since the theory parameter $\bar{\lambda}$ is fixed, condition (\ref{H0constraintmodelI}) determines the $\Omega_0$ value. Solid black line represents the standard flat $\Lambda$CDM model $H^2_{\Lambda CDM}=H^2_0\left[ 0.3 (1+z)^3 +0.7 \right]$. The dashed black line shows the pure matter dominated Einstein-de Sitter universe $H^2_{EdS}=H^2_0(1+z)^3$. We also show the expansion in model I for the values $\bar{\lambda}= -0.1$ (red) and $\bar{\lambda}= 1.5$ (blue). The EdS expansion is achieved with $\bar{\lambda}=0$. It is worth noting that one could promote a reasonable fit of data by adopting large $\bar{\lambda}$ values. However, such choice would lead to a pathological behavior for high redshifts. Therefore, a statistical analysis leading to the ``best-fit'' parameters would hide the inviability of the model as a whole. In order to assess the transition to the accelerated expansion phase we should calculate the deceleration parameter $q(z)=-1 - \dot{H}/H^2$. Calculating this quantity for model I,
\begin{equation}
q_I (z)=\frac{1}{2+3\bar{\lambda}},
\end{equation}
which is a constant value. Therefore, since the current acceleration expansion epoch is a well established model independent observational fact (see, e.g., \cite{Seikel:2008ms}), we conclude that as model I does not provide the transition from deceleration to a accelerated phase it should be ruled out.

Background expansion for model II is shown in the right panel of Fig. I. Again, by calculating the deceleration parameter in this case we find
\begin{equation}
q_{II}(z)= \frac{1+\bar{\gamma}_{2}\Omega}{2+5\bar{\gamma}_2 \Omega},
\end{equation}
which remains positive for all $\bar{\gamma}_2 > 0$ values. Therefore, we conclude once more that model II should also be ruled out as a viable alternative.

\begin{figure}[t]\label{lambda}
\includegraphics[width=0.4\textwidth]{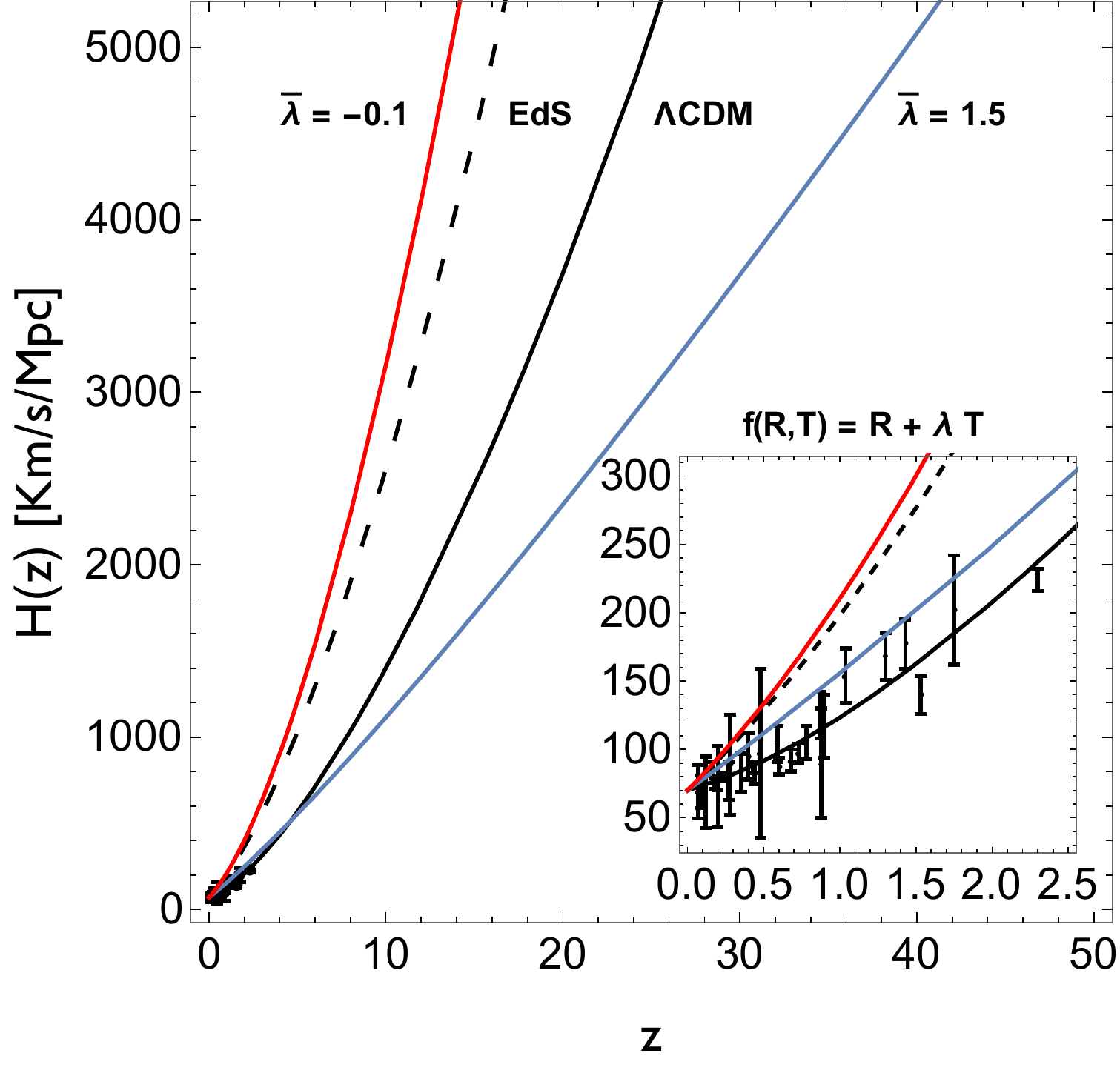}\hspace{1.5cm}
\includegraphics[width=0.4\textwidth]{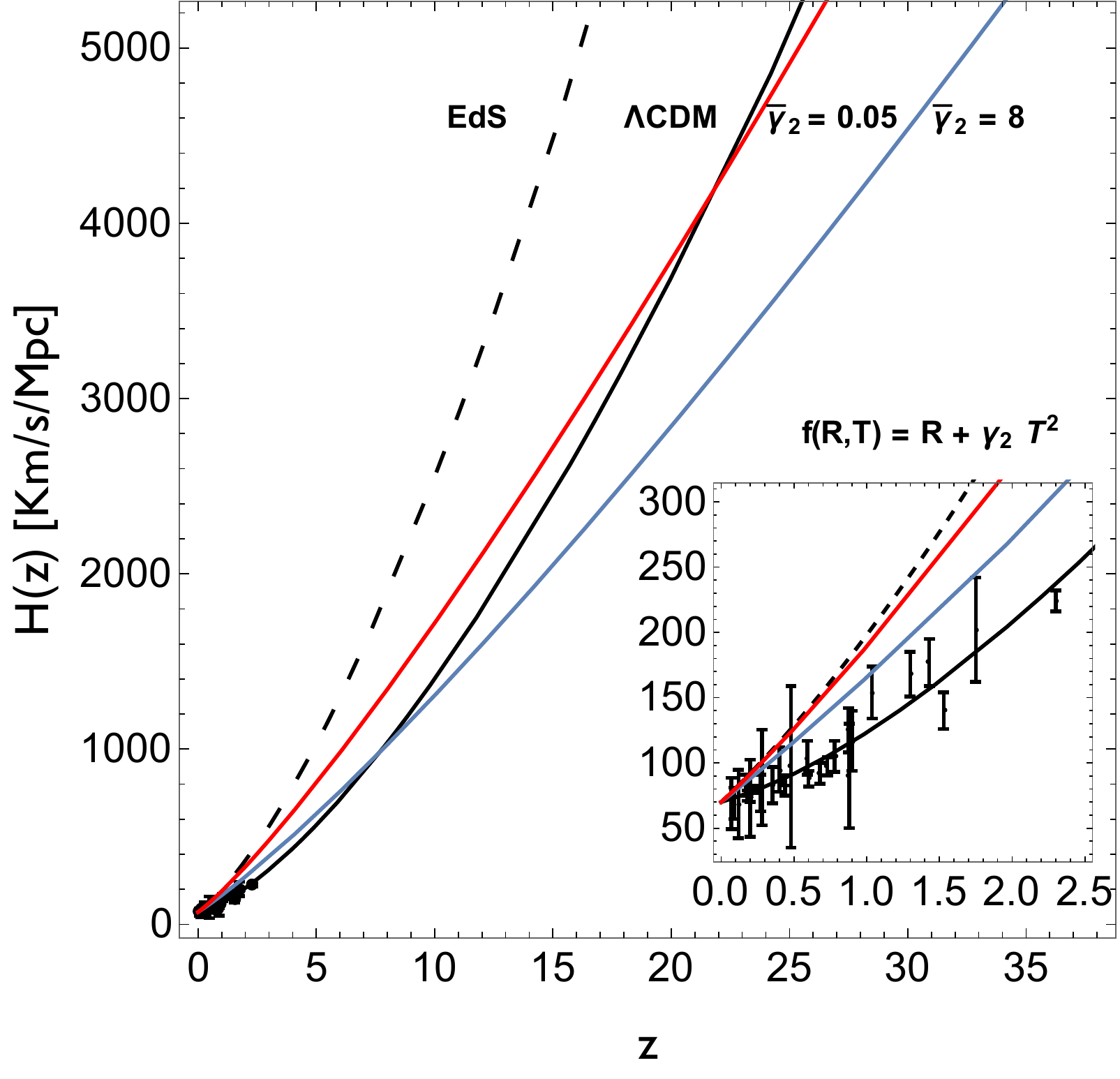}
\caption{\label{fig:i} Background expansion as a function of the redshift. The observational data plotted are obtained from the cosmic chronometers technique. The $\Lambda$CDM expansion is plotted in the solid-black line and the Einstein-de Sitter (Eds) expansion in dashed-black line. The inset in the figure amplifies the low-z region. Left: Model I. Right: Model II.}
\end{figure}


We focus now on model III and show the results in Fig. 2. In this case the usual conservation law for $\Omega$ applies, i.e., the right hand side of (\ref{conservation}) vanishes. Again, the Einstein-de Sitter evolution is recovered by setting $\bar{\gamma}_{1/2}=0$. Contrary to models I and II, the deceleration parameter in model III depends on time 
\begin{equation}
q_{III}(z)=\frac{2\Omega-\Omega^{1/2}\bar{\gamma}_{1/2}}{4(\Omega+\Omega^{1/2}\bar{\gamma}_{1/2})},
\end{equation}
and we have verified that the transition to acceleration occurs for any $\bar{\gamma}_{1/2} > 1.2$. According to (\ref{H0constraint}) the parameters are related via the expression
\begin{equation}\label{H0constraintmodelII}
1 = \Omega_0 +\bar{\gamma}_{1/2} \Omega^{1/2}_0 \frac{w+2}{2(1-3w)^{1/2}}.
\end{equation}

Since model III cannot be discarded immediately as the previous cases we proceed with a more robust data comparison using the Supernovae data. In the right panel of Fig. 2 we plot distance modulus $\mu$,
\begin{equation}
\mu = 5 {\rm log}_{10} \left(\frac{d_L}{10pc}\right) \hspace{1cm} {\rm with} \hspace{1cm} d_L = (1+z_{hel})c\int^{z_{cmb}}_{0}\frac{dz^{\prime}}{H(z^{\prime})},
\end{equation}
where $c$ is the speed of light and $z_{cmb}$ and $z_{hel}$ are respectively the CMB rest-frame and the heliocentric supernovae redshifts. We have also defined the luminosity distance $d_L$.
The data is taken from the Joint Light-Curve analysis (JLA) data set \cite{Betoule:2014frx} where the observed distance modulus is defined according to
\begin{equation}
\mu_{obs}=m^{\star}_{B}-M_B+\alpha \times X_1 - \beta \times \mathcal{C},
\end{equation}
where $m^{\star}_{B}$ is the $B$ band rest-frame observed peak magnitude, $\mathcal{C}$ describes de SN color at maximum brightness, $X_1$ describes the time strechting of the light-curve and the $M_B$ is the absolute $B-$band magnitude (see more details in \cite{Betoule:2014frx}). The parameters $\alpha$ and $\beta$ are free and should be determined with proper the statistical analysis. 

In both panels of Fig.2  we plot the $\Lambda$CDM model in the solid black line and the cases $\bar{\gamma}_{1/2}=2$ (red) and $\bar{\gamma}_{1/2}=3$ (blue). In the low-z region, the larger the parameter $\bar{\gamma}_{1/2}$, the closer model III is to the $\Lambda$CDM. The left panel shows on the other hand the contrary behavior on larger redshifts. 

We have also calculated the age of the universe in each configuration of model III and the results are compatible with the standard model. While the age of the universe is estimated as $13.7$ G$y$ in the $\Lambda$CDM model, we find $13.4$ G$y$ for $\bar{\gamma}_{1/2}$ and $14.5$ G$y$ for $\bar{\gamma}_{1/2}=3$.

\begin{figure}[t]\label{gamma05}
\includegraphics[width=0.4\textwidth]{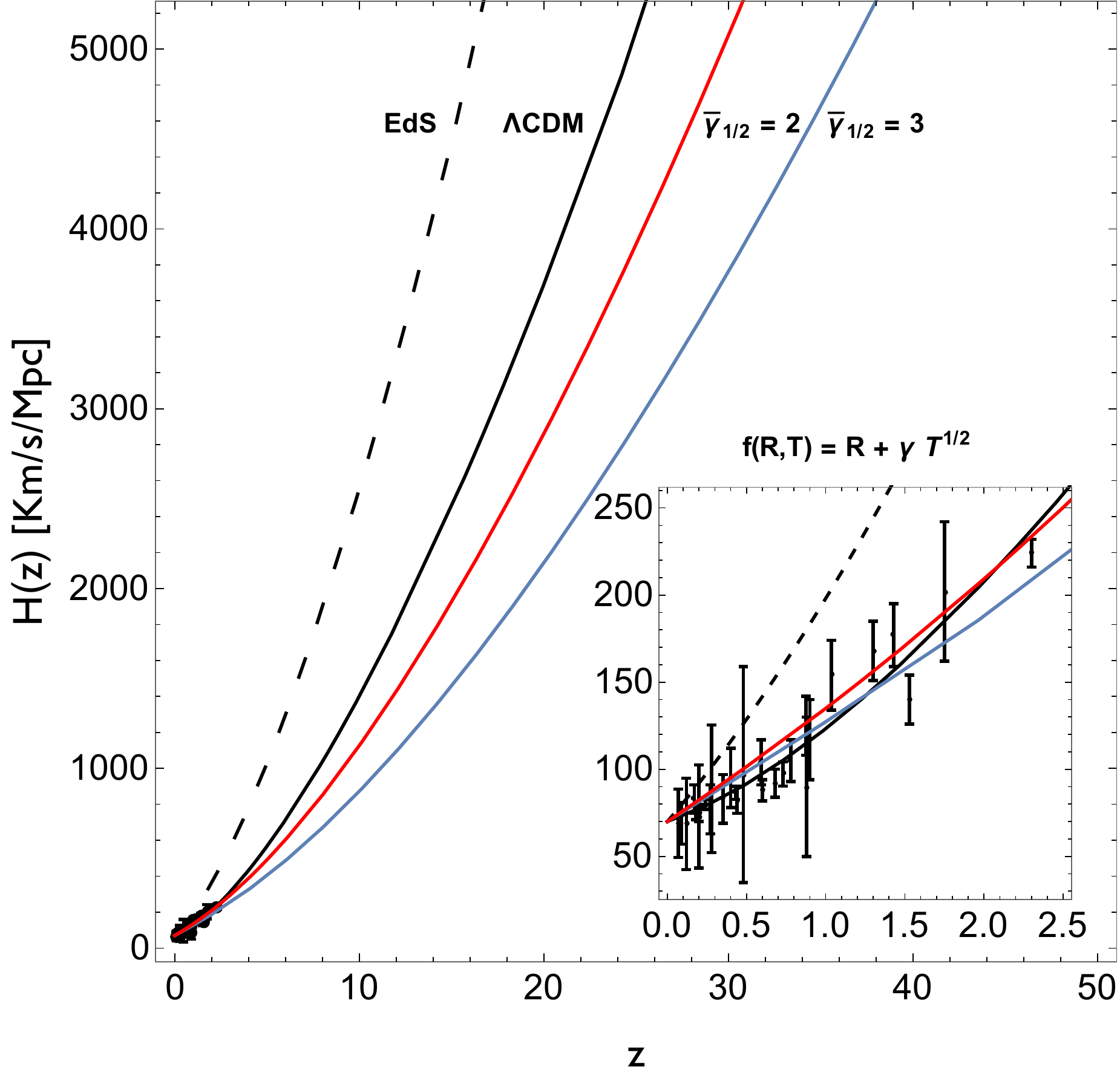}
\includegraphics[width=0.4\textwidth]{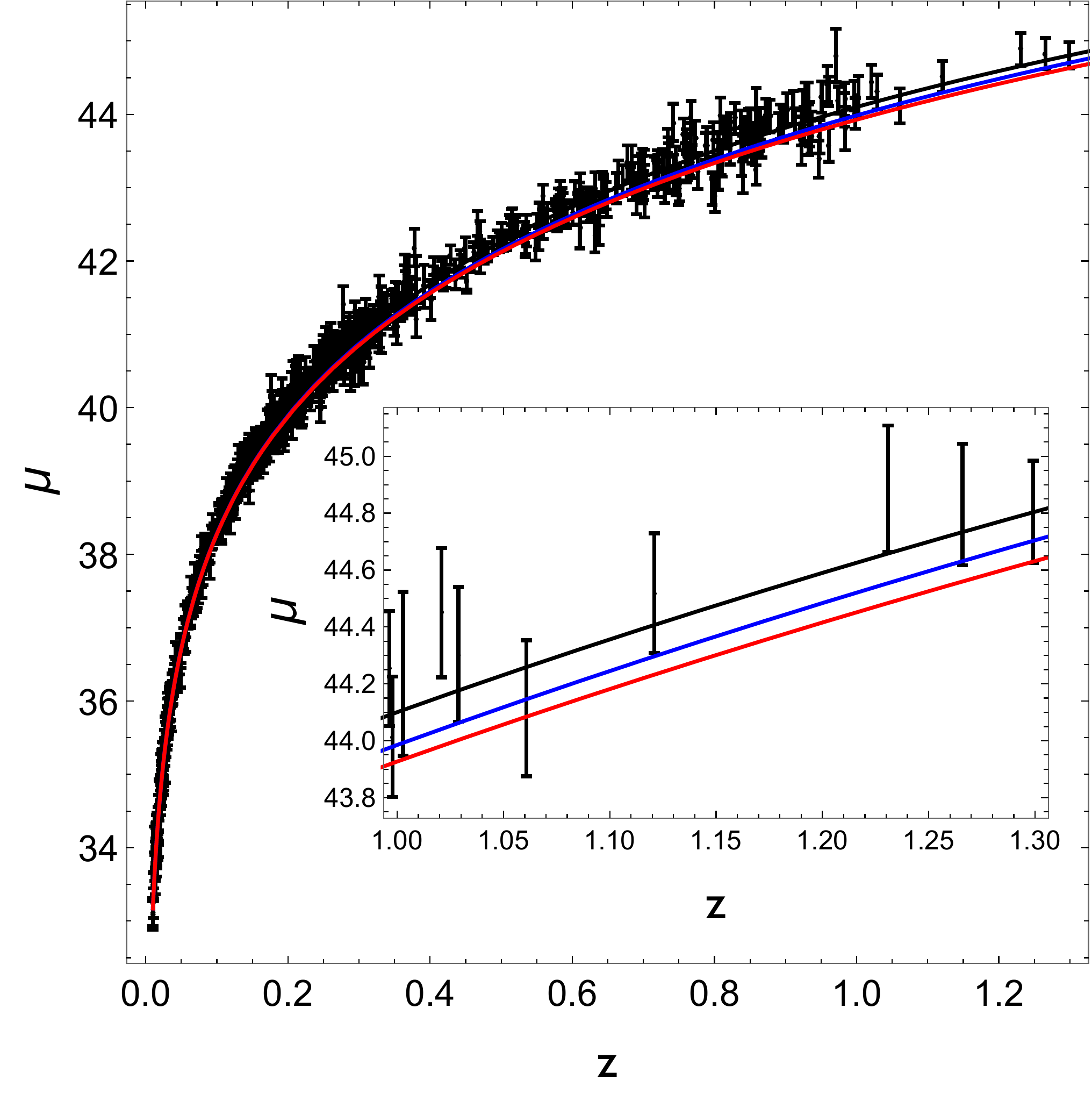}
\caption{\label{fig:i} Model III: In the left panel the Background expansion as a function of the redshift is shown. The observational data plotted are obtained from the cosmic chronometers technique. The $\Lambda$CDM expansion is plotted in the solid-black line and the Einstein-de Sitter (Eds) expansion in dashed-black line. Red line assumes $\bar{\gamma}_{1/2}=2$ and blue line assumes $\bar{\gamma}_{1/2}=3$. The inset in the figure amplifies the low-z region. In the right panel the distance modulus relation is shown with the JLA Supernovae data. The same colors apply in both panels.}
\end{figure}

Finally, we are to add a brief dicussion about the model IV, arguing why it in fact fails in providing an accelerated universe as we observe today, despite the claims raised in \cite{Moraes:2016jyi}. As we did for the previous cases, here we found the following deceleration parameter
\begin{equation}
\label{qIV}
q_{IV}(z) = \frac{1+\bar{\gamma}_2 \Omega}{2+3 \bar{\lambda}+5 \bar{\gamma}_2 \Omega}.
\end{equation}
Besides, for this model the constraint (\ref{H0constraint}) reduces to the form 
\begin{equation}
\label{constIV}
\frac{5}{2}\bar{\gamma}_{2}\Omega_{0}^{2}+\left(1+\frac{3\bar{\lambda}}{2}\right)\Omega_{0}-1=0.
\end{equation}
In \cite{Moraes:2016jyi} the authors show a transition from the decelerated to the accelerated stage when fixing properly the free parameters of the model. We can translate the model's parameters they used into the dimensionless ones (\ref{params}) adopted in our work. In doing so, we shall see that for $\bar{\lambda}$ this mapping leads to the choice $\bar{\lambda}=-2/3$. It is easy to notice that using this $\bar{\lambda}$'s value in (\ref{constIV}) one brings up the relation $\Omega_{0}=\sqrt{\frac{2}{5\bar{\gamma}_2}}$, which obliges the $\bar{\gamma}_2$'s sign to be positive, as the density parameter today is obviously a real (and positive) number. Substituting $\bar{\lambda}=-2/3$ in (\ref{qIV}) we are left with a new deceleration parameter in the form
\begin{equation}
\label{qIV0}
\tilde{q}_{IV}(z)=\frac{1+\bar{\gamma}_2 \Omega}{5 \bar{\gamma}_2 \Omega}. 
\end{equation}
Since $\Omega(z)$ as well as the model parameter $\bar{\gamma}_2$ have to be positive, the $\tilde{q}_{IV}(z)$ must be also positive for all $z$, what describes a decelerated universe which cannot transit to a accelerated phase. This result contradicts what the authors have found in \cite{Moraes:2016jyi}, where they assert to have achieved such transition for the model IV. However, we noticed that such conclusion is consequence of a mistake in the evolution equation for the scale factor, what makes them ending up with a wrong solution for $a(t)$ and hence an incorrect deceleration parameter. 

In the Appendix \ref{A} we can see in more detail how this mistake have constituted a crucial factor for the result they obtained. Besides, we show that by fixing this error and computing the correct deceleration parameter, we get a cosmology in which the universe is unable to transit from a decelerated to a accelerated stage, confirming what is expected from (\ref{qIV0}) as discussed in the previous paragraph.

\section{$F(R,T)$ as an alternative for dark matter } 

Recently, Ref. \cite{Zaregonbadi:2016xna} investigated $f(R,T)$ theories in the context of galactic rotation curves. Then, this is a clear attempt to solve the dark matter problem via a modified gravity theory. Following such perspective we ask now whether $f(R,T)$ theories can also afford an explanation for the cosmological dark matter effect on the background expansion. Let us consider a late time cosmological model with its total energy momentum tensor composed uniquely by the known pressureless baryonic component i.e., $T^{\mu\nu}=T^{\mu\nu}_b$. The accelerated expansion is therefore achieved by including a $\Lambda$-like effect on the expansion with the help of the choice $n=0$. Then, expansion \ref{H1} is rewritten as

\begin{equation}\label{H2}
\frac{H^2}{H_0^2} = \Omega_b + \frac{3}{2}\bar{\lambda}_0\Omega_b+\frac{\bar{\gamma}_0}{2}.
\end{equation}
Notice that $\bar{\gamma}_0$ plays the role of the cosmological constant. The conservation law for the baryonic energy density obeys to
\begin{equation}
\dot{\Omega}_b+3H\Omega_b=\frac{-\bar{\lambda}_0}{2(1+\bar{\lambda}_0)}\dot{\Omega}_b.
\end{equation}
Keeping the conservative side, we also assume that the baryonic sector is constrained by BBN results leading to a todays's fractionary density $\Omega_{b0}=0.04$.

The deceleration parameter in this scenario is written as
\begin{equation}
q(z) = \frac{1}{2}\frac{1-\frac{\bar{\gamma}}{\Omega_b}}{1+\frac{3\bar{\lambda}_0}{2}+\frac{\bar{\gamma}}{2\Omega_b}},
\end{equation}
and it does lead to a transition to the accelerated phase.

In Fig.3 we plot the background expansion in this case. It is worth noting the disagreement with the expected pure dust like behavior $H \sim (1+z)^{3/2}$ on large redshifts. 

\begin{figure}[t]\label{gamma05}
\includegraphics[width=0.4\textwidth]{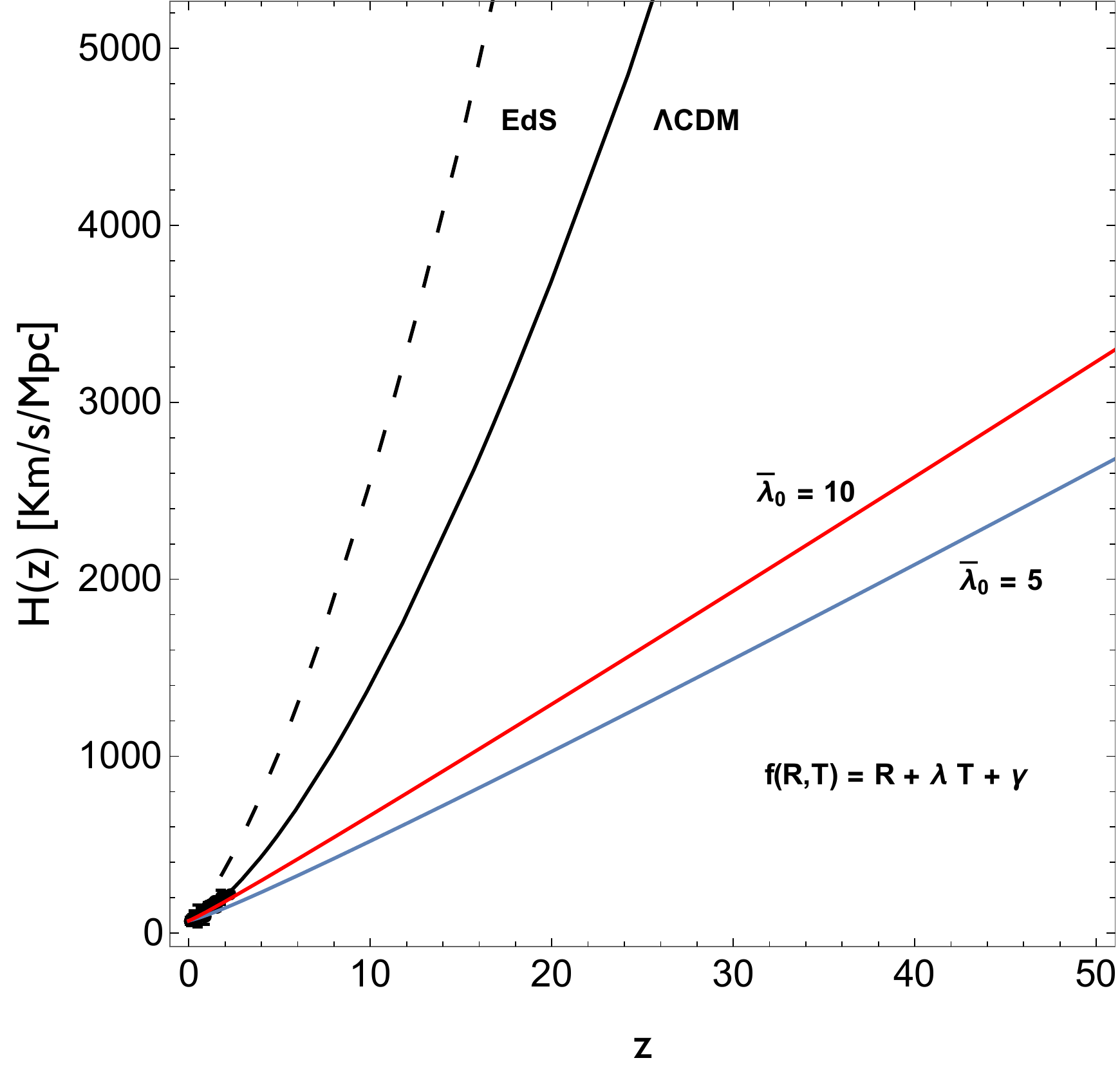}\hspace{0.5cm}
\includegraphics[width=0.4\textwidth]{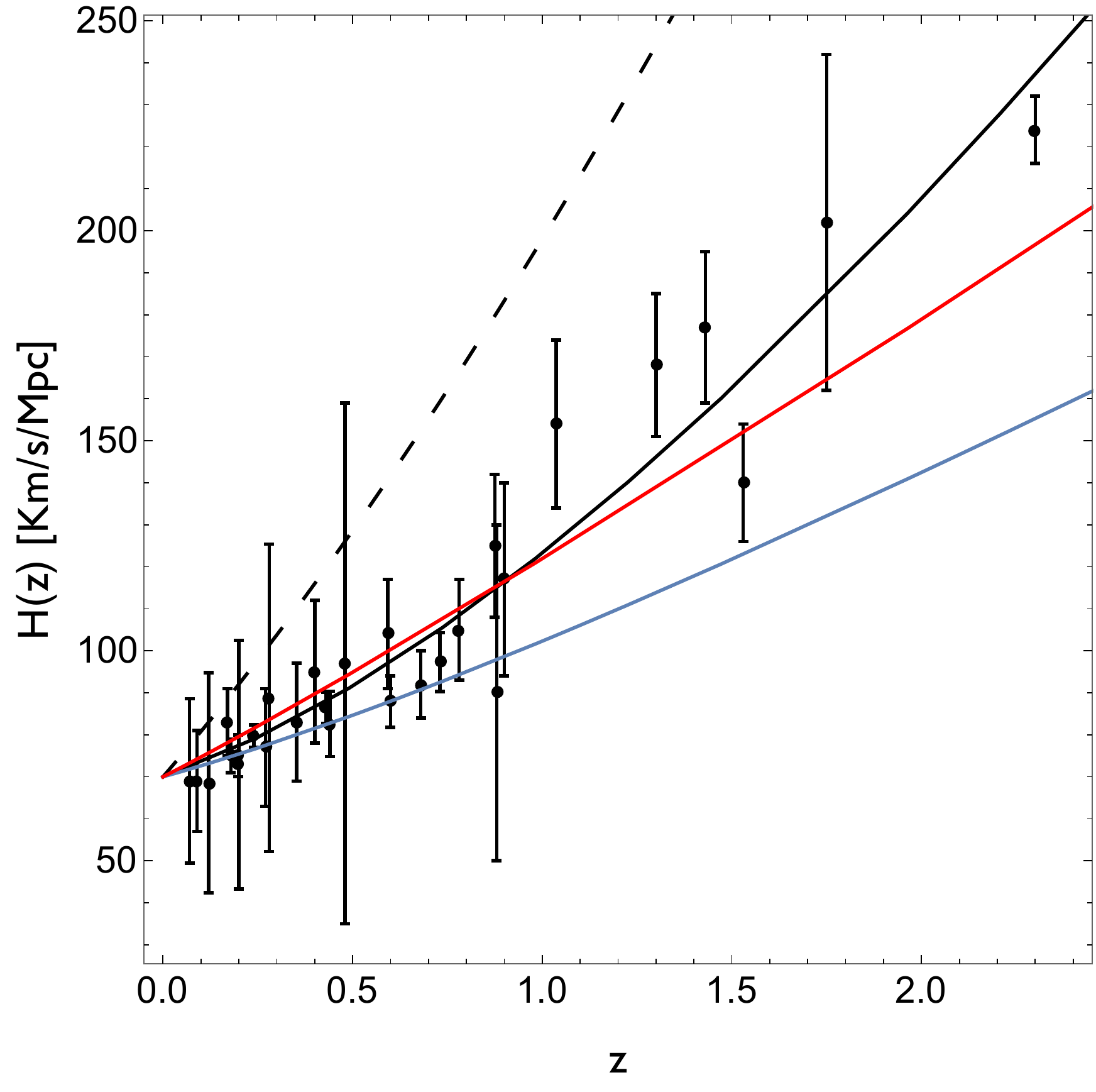}
\caption{\label{fig:i} The Background expansion as a function of the redshift is shown for the case $F(R,T)$ replaces the dark matter sector, Eq.  (\ref{H2}) . The observational data plotted are obtained from the cosmic chronometers technique. The $\Lambda$CDM expansion is plotted in the solid-black line and the Einstein-de Sitter (Eds) expansion in dashed-black line. Red line assumes $\bar{\lambda}_{0}=10$ and blue line assumes $\bar{\lambda}_{0}=5$. The right panel amplifies the low-z region.}
\end{figure}

\section{Final Discussion}

In this paper the viability of the $f(R,T)$ models in the cosmological realm was discussed. We employed such theories to address the two main dilemmas of the modern cosmology - the dark energy and dark matter. As such, we replaced such dark components by $f(R,T)$-like modifications of the standard gravity. Among many possible shapes for $f(R,T)$ we selected those ones which are often used in the literature whose forms can be encoded in the choice $f(R,T)= R + \lambda T + \gamma_{n} T^n$. This choice branches off into the models I, II, III and IV presented in the third section of this article. The model III ($\lambda=0$ and $n=1/2$) is particularly interesting as it is the only one which obeys the usual continuity equation. Aiming at providing a description for the cosmological background consistent with the observed cosmic acceleration we have confronted these models with $H(z)$ data, and for model III we have reinforced our analysis by making a comparison with Supernovae data. While the $H(z)$ analysis shows a visible consistency of the models I and II with $\Lambda$CDM at low redshifts, it is noticed a significant deviation of these models from the standard scenario as one goes back in the past, evidencing clear obstacles faced by such class of $f(R,T)$ theories in describing correctly the cosmological evolution as a whole. Besides, these difficulties become more evident when we compute the deceleration parameter for either model: the both show up unviable as they lead to universes without a transition from a decelerated stage to a accelerated one, as it is expected for any cosmology compatible with the observational predictions. We consider that this is enough to rule out these models as viable scenarios for cosmology. On the other hand, the model III do manifests such deceleration-acceleration transition, coming up as a possible route towards a realistic cosmological picture. 
Insisting on the model III study, we have verified that its disagreement with both the $H(z)$ and Supernovae data increases at higher redshift regimes, whereas it is less notable at lower redshifts. However, even this case presents an inconvenient behavior when compared with the data: whilst at very low-z this model approaches the $\Lambda$CDM for increasing values of the coefficients $\bar{\gamma}_{1/2}$, this pattern reverses at high-z ($z\gtrsim 2$) and the model get closer and closer to the standard cosmology as the parameters $\bar{\gamma}_{1/2}$ assume lower and lower numerical values. Besides, using the Supernovae dataset one observes a slight departure from the data for larger redshifts. These incongruencies indicate a serious disadvantage of the model III, showing its failure in providing a consistent description of the background dynamics. Furthermore, for the model IV we have shown in details that the supposed deceleration-acceleration transition claimed in the literature is an incorrect result. 

Then, we have also investigated the viability of an alternative hypothesis, according which the dark matter is interpreted as a natural consequence of a $f(R,T)$ theory, while the cosmic speed-up keeps being driven by a cosmological constant. We have concentrated on the case $n=0$ in (\ref{fRT}) and verified that for such assumption the $H(z)$ evolution shows a strong contradiction with the corresponding data set when one looks at larger redshifts. {Our analysis challenges the $f(R,T)$ explanations of the DM phenomena. But this result applies only to the particular way we have conceived equation (\ref{H2}).}

In fact, we have pointed out in this work difficulties that $f(R,T)$ gravity present when one tries to realise usual cosmic expansion. It was possible to build up new models either by adding extra fields (e.g. radiation, neutrinos or scalar fields) or other types of $f(R,T)$ functions and test the viability of them.

Also, concerning the dark matter description via $f(R,T)$ gravity, it is important to realise that the background expansion does not reach an Einstein-de Sitter evolution on large redshifts. Then, the standard matter dominated expansion is not achieved and structure formation can be affected. However, results from Ref. \cite{alvarenga} indicate a tendency for a wavenumber dependent super matter agglomeration for $f(R,T)$ theories when compared to the standard $\Lambda$CDM scenario i.e., the nonlinear stage is reached much earlier in the $f(R,T)$ case. Perhaps, this trend can be compensated by the faster background expansion ($z>>1$) found in Fig. 3 leading to a viable scenario. A careful analysis is demanded here and we leave this investigation for a future work.

Finally, it is worth noting that the approach we have employed in this work can be immediately extended to similar theories like $F(R, T_{\phi})$ \cite{Harko:2011kv} and $F (R, T, R_{\mu\nu} T^{\mu\nu} )$ \cite{OdintsovSaez}. This analysis should appear in future works.

\noindent
{\bf Acknowledgement:} We thank CNPq (Brazil) and FAPES (Brazil) for partial financial support. We thank P. Moraes, J. Fabris, S. Nojiri, M.E. Rodrigues, D. Sa\'ez-Gom\'es, S. Odintsov and F. Sbis\'a for much appreciated discussions.

\appendix
\section{Computing the correct deceleration parameter for $\bar{\lambda}=-2/3$}\label{A}
For sake of comparison with the reference \cite{Moraes:2016jyi}, let us express the model IV as $f(R,T)=R+\alpha T + \beta T^2$. For this theory, considering $p=0$, it is straightforward to see that the equations (\ref{eq00}) and (\ref{eq11}) turn out to be
\begin{equation}
\label{eqq00}
3\left(\frac{\dot{a}}{a}\right)^2=8 \pi G \rho +\frac{1}{2}\left(3 \alpha+5\beta \rho\right)\rho,
\end{equation}
and
\begin{equation}
\label{eqq11}
2\frac{\ddot{a}}{a}+\left(\frac{\dot{a}}{a}\right)^2=\frac{1}{2}\left(\alpha+\beta\right)\rho,  
\end{equation}
which coincide with the equations (4) and (5) of \cite{Moraes:2016jyi} for dust. Then, let us resort to the same fixing they assume for this model in which $\alpha=-\frac{16\pi G}{3}$, what leads the equation above to a simpler form
\begin{equation}
\label{eqq0}
3\left(\frac{\dot{a}}{a}\right)^2=\frac{5 \beta}{2}\rho^2,
\end{equation}
and
\begin{equation}
\label{eqq1}
\frac{\ddot{a}}{a}+\frac{1}{5}\left(\frac{\dot{a}}{a}\right)^2=-\frac{4\pi G}{3}\rho.
\end{equation}
The next step is to employ (\ref{eqq0}) to write $\rho$ in terms of the scale factor and its first time derivative as $\rho=\sqrt{\frac{6}{5\beta}}\left(\frac{\dot{a}}{a}\right)$. Then using it in (\ref{eqq1}) we obtain
\begin{equation}
\label{edat}
\frac{\ddot{a}}{a}+\frac{1}{5}\left(\frac{\dot{a}}{a}\right)^2=-\frac{4\pi G}{3}\sqrt{\frac{6}{5\beta}}\left(\frac{\dot{a}}{a}\right).
\end{equation}
Let us notice that this equation is different from that one obtained in \cite{Moraes:2016jyi}\footnote{See eq. (7) of that reference which seems to be incorrect. Moreover, let us notice that this equation shows a summation between two terms with incompatible units.}. By multiplying (\ref{edat}) by $a/\dot{a}$, we can express this equation as follows
\begin{equation}
\label{edat1}
\frac{d}{dt}\left(\ln \dot{a} + \frac{1}{5}\ln a\right)=-\alpha_{1},
\end{equation}
where we have defined $\alpha_{1}\equiv\frac{4\pi G}{3}\sqrt{\frac{6}{5\beta}}$. Integrating (\ref{edat1}) we find the following solution
\begin{equation}
\label{solat}
a(t)=a_0(e^{-\alpha_1 t}-1)^{5/6}.
\end{equation}
Here we have fixed one of the integration constants by assuming the initial condition $a(0)=0$. As expected we obtained a distinct scale factor which by turn shall yield a different deceleration parameter 
\begin{equation}
\label{qIV1}
\hat{q}_{IV}(t)=-1+\frac{6}{5}e^{\alpha_1 t}.
\end{equation}
Since the argument of the exponential factor is always positive, the second term on the left hand side shall keep greater than $1$, for every value of $\alpha_1$ and for every instant of time. Thus, no deceleration-acceleration transition is observed for this case, contrarily to the authors in \cite{Moraes:2016jyi} claimed, but in accordance with we argue in the section III.


\begin{thebibliography}{} 
\bibitem{MG} S. Tsujikawa, Lect.Notes Phys.{\bf 800}, 99 (2010);  S. Capozziello, M. de Laurentis, Phys. Rep. {\bf509} 167 (2011); T. Clifton, P. G. Ferreira, A. Padilla, C. Skordis, Physics Reports {\bf513}, 1 (2012); Shin'ichi Nojiri and Sergei D. Odintsov, Phys.Rept. {\bf505},  59(2011); E. Berti, et al, Classical Quant. Grav. {\bf32}, 243001 (2015). 




\bibitem{Harko:2011kv} 
  T.~Harko, F.~S.~N.~Lobo, S.~Nojiri and S.~D.~Odintsov,
  Phys.\ Rev.\ D {\bf 84}, 024020 (2011).

\bibitem{Momeni:2011am} 
  M.~Jamil, D.~Momeni, M.~Raza and R.~Myrzakulov,
  Eur.\ Phys.\ J.\ C {\bf 72}, 1999 (2012).

\bibitem{Seikel:2008ms} 
  M.~Seikel and D.~J.~Schwarz,
  JCAP {\bf 0902}, 024 (2009).

\bibitem{alvarenga} 
  F.~G.~Alvarenga, A.~de la Cruz-Dombriz, M.~J.~S.~Houndjo, M.~E.~Rodrigues and D.~Sáez-Gómez,
  Phys.\ Rev.\ D {\bf 87}, no. 10, 103526 (2013);
  Erratum: [Phys.\ Rev.\ D {\bf 87}, no. 12, 129905 (2013)].


\bibitem{misner}
 C.~W.~Misner, K.~S~.Thorne and J.~A.~Wheeler, {\it Gravitation}, San Francisco: Freeman, (1973).
 
\bibitem{wald} R. Wald, {\it General Relativity}, University of Chicago Press, Chicago, (1984).

\bibitem{schutz} B. Schutz, {\it A first course in General Relativity}, Cambridge University Press, Cambridge, United Kingdom, (1985).

\bibitem{dinverno} R. D'Inverno, {\it Introducing Einstein's Relativity}, Clarendon Press, Oxford, England, (1992).


\bibitem{weinberg} S. Weinberg, {\it Gravitation and cosmology: principles and applications of the general theory of relativity}, John Wiley \& Sons, Inc.;(1972).

\bibitem{Moraes:2016jyi} 
  P.~H.~R.~S.~Moraes, G.~Ribeiro and R.~A.~C.~Correa,
  Astrophys.\ Space Sci.\  {\bf 361}, no. 7, 227 (2016).

\bibitem{BarrientosO.:2014ska} 
  J.~Barrientos O. and G.~F.~Rubilar,
  Phys.\ Rev.\ D {\bf 90}, no. 2, 028501 (2014).

\bibitem{Baffou:2013dpa} 
  E.~H.~Baffou, A.~V.~Kpadonou, M.~E.~Rodrigues, M.~J.~S.~Houndjo and J.~Tossa,
  Astrophys.\ Space Sci.\  {\bf 356}, no. 1, 173 (2015)
  doi:10.1007/s10509-014-2197-z
  [arXiv:1312.7311 [gr-qc]].

\bibitem{Momeni:2015fyt} 
  D.~Momeni, P.~H.~R.~S.~Moraes and R.~Myrzakulov,
  Astrophys.\ Space Sci.\  {\bf 361}, no. 7, 228 (2016).

\bibitem{Moraes:2016mlp} 
  P.~H.~R.~S.~Moraes and R.~A.~C.~Correa,
  arXiv:1606.07045 [gr-qc].
	
\bibitem{Zaregonbadi:2016xna} 
  R.~Zaregonbadi, M.~Farhoudi and N.~Riazi,
  Phys.\ Rev.\ D {\bf 94}, 084052 (2016).
	
	
	\bibitem{Hz} R.  Jimenez  and  A.  Loeb,
Constraining  Cosmological  Pa-  rameters  Based  on  Relative  Galaxy  Ages
, Astrophys.J.
573 (2002) 37; R. Jimenez, L. Verde, T. Treu and D. Stern,
Constraints on the equation of sate of Dark Energy and the
Hubble Constant from Stellar Ages and the Cosmic Microwave Background
, Astrophys. J. 593 (2003) 622; D.  Stern,  R.  Jimenez,  L.  Verde,  M.  Kamionkowski  and  S.  Adam  Stanford,
Cosmic  Chronometers: Constraining the Equation of State of Dark Energy. I: H(z) Measurements
, JCAP 1002 (2010) 008.
	
\bibitem{Hzdata} O. Farooq, D. Mania and B. Ratra,
Hubble parameter measurement constraints on dark energy
, Astroph. J., 764 (2013) 138. M. Moresco et al., New  constraints  on  cosmological  parameters  and  neutrino  properties  using  the  expansion rate of the Universe to z 1.75, JCAP 07 (2012) 053.

\bibitem{Zaregonbadi:2016xna} 
  R.~Zaregonbadi, M.~Farhoudi and N.~Riazi,
  Phys.\ Rev.\ D {\bf 94}, 084052 (2016).
	
\bibitem{Betoule:2014frx} 
  M.~Betoule {\it et al.} [SDSS Collaboration],
  Astron.\ Astrophys.\  {\bf 568}, A22 (2014).
	

\bibitem{OdintsovSaez}
Sergei D. Odintsov and Diego Sáez-Gómez, Phys.Lett. {\bf B725}, 437 (2013).
	
	
	
	
\end{thebibliography}
\end{document}